\documentclass[prl,twocolumn]{revtex4}
\usepackage{graphicx}
\usepackage{dcolumn}
\usepackage{multirow}
\usepackage{amsmath,amsthm,amssymb}
\usepackage{subfigure}

\def\ep{\varepsilon}

\begin{document}
\title{A comment on "Boson stars in AdS"}

\author{Maciej Maliborski}
\email{maliborski@th.if.uj.edu.pl}
\affiliation{M. Smoluchowski Institute of Physics, Jagiellonian University, Krak\'ow, Poland}

\author{Andrzej Rostworowski}
\email{arostwor@th.if.uj.edu.pl}
\affiliation{M. Smoluchowski Institute of Physics, Jagiellonian University, Krak\'ow, Poland}
\date{\today}
\begin{abstract}
We comment on asymptotically AdS "large-$\sigma$" initial data,
studied in Phys. Rev. \textbf{D87}, 123006 (2013), that are immune to
turbulent instability.
\end{abstract}

\maketitle

\begin{figure}[h]
\includegraphics[width=0.4\textwidth]{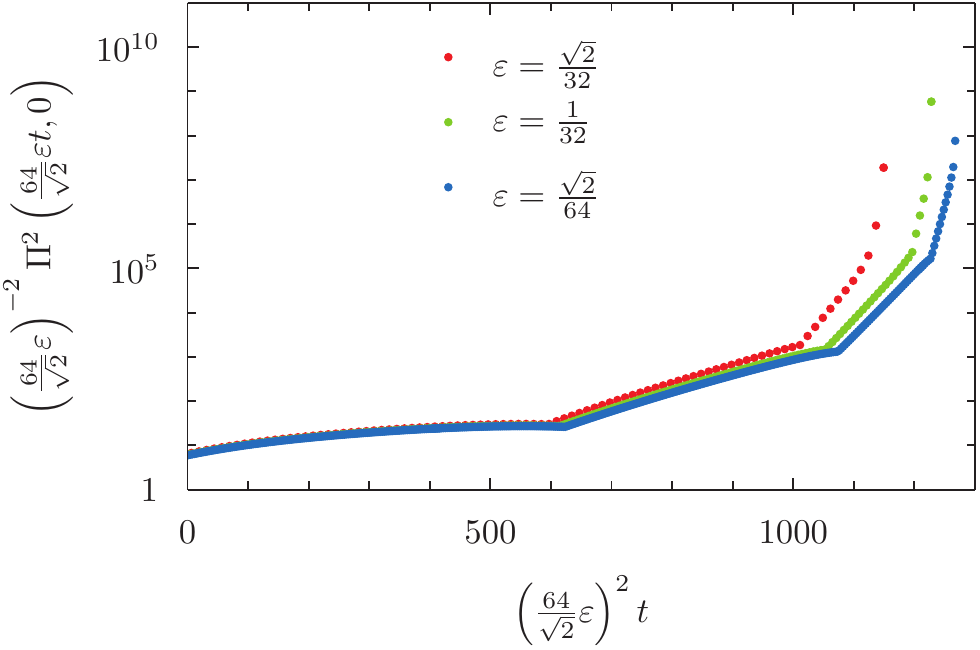}
\caption{The rescaled value of $\Pi^2(t,0)$ for solutions with
  initial data \eqref{ID} for three moderately small amplitudes and
  $\sigma=8$. For clarity of the plot only the upper envelopes of
  rapid oscillations are depicted. After making between about a ninety
  (for $\ep=\sqrt{2}/32$) and four hundred (for $\ep=\sqrt{2}/64$)
  reflections, the solutions finally collapse.}
\label{fig1}
\end{figure}

In a recent paper \cite{bll2} Buchel, Liebling and Lehner (BLL) showed
that boson stars in AdS${}_{3+1}$ are non-linearly stable. This is
important result that, together with \cite{mr}, substantiates the
claims of \cite{dhms} that there exists a class of stable,
asymptotically AdS solutions of Einstein equations. As a by product,
in sec. 4 of their paper, BLL also study Einstein-AdS -- massless
scalar field system investigated in \cite{br}. In particular, they
show
that the family of initial data
\begin{equation}
\label{ID}
\Phi(0,x)=0, \qquad \Pi(0,x) = \frac {2 \ep} {\pi}\,\exp \left( -
\frac{4 \tan^2 x}{\pi^2 \sigma^2} \right)
\end{equation}
used in \cite{br} to conjecture the turbulent instability of AdS is in
fact immune to this instability for the values of $\sigma \sim
0.5$. Then, they extrapolate this result to $\sigma \gtrsim 0.5$ and
speculate that the fact that \textit{"the large-$\sigma$ initial data
  is immune to the weakly turbulent instability is consistent with the
  argument that widely distributed mass energy prevents the coherent
  amplification typical of the instability"}. While we confirm their
numerical result, we disagree with its interpretation. We think that
the correct explanation of the observed immunity against instability
comes from the fact that in $3+1$ dimensions, the initial data
(\ref{ID}) with $\sigma\approx 1$ and small $\ep$ belong to the
stability island of the time-periodic solution bifurcating from the
fundamental mode of the massless scalar field propagating on a fixed
AdS background \cite {mr}.
We support our claim with the following observations:
\begin{enumerate} 
\item It turns out that BLL's extrapolation to $\mbox{$\sigma \gtrsim 0.5$}$ is
  unjustified, namely for still larger values of $\mbox{$\sigma \gtrsim 8$}$ the
  turbulent cascade is restored as depicted in Fig.~\ref{fig1}
  (analogous to Fig.~2 of \cite{br}).
\item We have not found such immunity to instability for the initial
  data (\ref{ID}) in higher spatial dimensions ($d>3$). Of course, we
  can not exclude that one can fine tune $\sigma$ to achieve that but
  that does not seem to be as straightforward as in $3+1$. It is
  consistent with the fact that the radius of convergence of
  perturbative series \cite{mr} decreases as we go to higher
  dimensions and/or mode numbers. Thus, we expect the stability
  islands of higher modes (in dimension and/or wave number) to shrink.
\item If we Fourier transform $\hat \Pi(\omega,0) = \int \Pi(t,0)
  e^{-i \omega t}\,dt$ we find that for the values of parameters in
  (\ref{ID}) corresponding to the stability window, $\Pi(\omega,0)$ is
  peaked around $\omega=3$, corresponding to the frequency of a time
  periodic solution bifurcating from the fundamental mode. For example
  for $(\sigma,\, \ep)=(1/2, \, 2)$ we get a dominant peak in the
  spectrum at $\omega \approx 3.15$. 
\end{enumerate}
The results of \cite{br,bll2,dhms,mr} suggest that while AdS solution
itself is unstable against black hole formation, putting some
fine-tuned small ripples on AdS in the form of time-periodic solutions
or boson stars prevents the instability. Thus the dynamics of
asymptotically AdS spacetimes turns out to be much more complicated
that could have been expected, and probably for still some time the
combined numerical/symbolic
approach will be the basic tool for studying this problem.

\noindent \emph{Acknowledgments:} We are indebted to
Alex Buchel, Steven L. Liebling, Luis Lehner and Piotr Bizo\'n for discussion.

\end{document}